# Possible High-Tc Superconductivity in LiMgN: A MgB$_2$-like Material

O. P. Isikaku-Ironkwe[1, 2]
[1]The Center for Superconductivity Technologies (TCST)
Department of Physics,
Michael Okpara University of Agriculture, Umudike (MOUAU),
Umuahia, Abia State, Nigeria
and
[2]RTS Technologies, San Diego, CA 92122

## Abstract

The search for superconductivity in materials iso-structural and iso-valent with magnesium diboride, MgB$_2$, has not yielded any results close enough to the 39K Tc of MgB$_2$. Lithium magnesium nitride, LiMgN, resembles MgB$_2$ in that they both have the same averages of electronegativity, valence electron count and atomic number. Their formula weights are very close too. Using our recently published chemical symmetry rules for estimating superconductivity and Tcs of materials, we predict that LiMgN, a semiconductor at room temperature, with the same material specific characterization dataset (MSCD) as MgB$_2$, should be a superconductor, in the ordered α-phase or the orthorhombic phase, when cooled to 39K.

## Introduction

The surprising discovery of high-Tc superconductivity at 39K in MgB$_2$[1, 2] in 2001 excited expectations that there may be other simple and similar MgB$_2$-like binary or ternary superconductors with the same electronic, structural and Tc similarity. Predictions and searches, based on iso-valent, iso-structural and DFT calculations have been undertaken both theoretically [3, 4, 5, 6] and experimentally [7, 8, 9, 10, 11, 12, 13, 14, 15, 16], without producing superconductors with Tcs anywhere near 39K. Many chemical-based reasons have been advanced too on diborides superconductivity [32] but none has produced or predicted correctly new MgB$_2$-like superconductors with Tc near 39K.

In 2007, we decided to look at this problem from slightly different parameters. We initiated studies [17, 18, 19], based on empirical material specific correlations of Tc with averages of electronegativity, valence electrons, formula weight and atomic number. One of the many materials we identified as a possible MgB$_2$-like superconductor, based on these studies is LiMgN. Further studies revealed the symmetry rules governing superconductors with similar electronegativity, valence electrons, atomic number and formula weight. Guided by these symmetry rules [20] we have been able to identify, using the Periodic Table only, and without recourse to DFT calculations, many potential superconductors. We also showed how to estimate Tc using the symmetry methods [20] and predicted 21 potential superconductors. This paper starts by reviewing the structural and electronic properties of LiMgN and methods for its preparation. Next we present the basis for making our prediction. We next show why iso-structural and iso-valent features are not sufficient conditions for superconductivity. We conclude that LiMgN should be a superconductor, based on the simple chemical symmetry rules.

## Preparation of LiMgN

Many routes have been used in the preparation of LiMgN. The preparation of LiMgN from precursor LiMg annealed in pure nitrogen at 800 degrees centigrade for 8 hours is well described in reference [21]. The orthorhombic structure type has been prepared from $Li_3N$ and $Mg_3N_2$ by rapid cooling from 1000K to room temperature [22]. LiMgN can be also be prepared by a hydriding thermal reaction between $MgH_2$ and $LiNH_2$ around 80 Kbar $H_2$ pressure and 723K and a subsequent dehydrogenation reaction [23] given by:

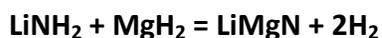
$$LiNH_2 + MgH_2 = LiMgN + 2H_2$$

Preparation of LiMgN as a commercial safe hydrogen storage compound is described in [24].

## Properties of LiMgN

LiMgN is a member of the filled tetrahedral semiconductor family LiMgX(X= N, P, and As) with a zincblende-like structure [21]. It can form in any of its two cubic phases known as ordered (α) and disordered (β), with the α- LiMgN more stable, with an experimental energy band gap of 3.2eV and cubic lattice constant of 4.995 [21]. LiMgN with an orthorhombic structure and lattice constants, a =7.1586Å, b = 3.5069Å, and c = 5.0142Å, and space group Pnma, has also been identified [22]. It has a cation-ordered antifluorite-type structure.

The electronic band structure of LiMgN has been extensively studied by a*b initio* DFT methods [25, 26, 27, 28] and also experimentally [21, 22, 28]. One study [25] indicates very strong covalent bonding of Li-N and Mg-N in LiMgN while the others [26, 27, 28] suggest strong ionic bonding of Li-N and Mg-N in disordered β-LiMgN. Raman spectrum studies [28] confirm antiflorite structure with space group Fm3m and that LiMgN is a filled tetrahedral semiconductor with band-gap of 3.3eV.

**Symmetry Rules: Framework for Predictions**
We showed [20] that the maximum Tc of a material may be expressed in material specific parameters of electronegativity, $\mathcal{X}$, valence electrons, Ne, and atomic number, Z, given by

$$\mathbf{T_c} = \mathcal{X} \frac{\mathrm{Ne}}{\sqrt{Z}} \mathrm{K_o} \tag{1}$$

where $\mathrm{K_o}$ is a parameter that determines the value to Tc. Ko = n(Fw/Z) and n is dependent on the family of superconductors. Fw represents formula weight of the superconductor. For MgB$_2$, Ko = 22.85 and Fw/Z is 6.26, making n = 3.65. Recently [20] we proposed that similar superconductors may arise when we compare their averages of electronegativity, $\mathcal{X}$, valence electrons, Ne, and atomic number, Z, and Fw/Z. We distinguished four possible cases, when at least two of the features are the same, namely: (a) $\langle \mathcal{X}, \mathrm{Ne}, \mathrm{Z}\rangle$, (b) $\langle \mathrm{Ne}, \mathrm{Z}\rangle$, (c) $\langle \mathcal{X}, \mathrm{Ne}\rangle$ and (d) $\langle \mathcal{X}, \mathrm{Z}\rangle$. We described such superconductors as similar. We found the symmetry rules apply within the range 0.75< Ne/$\sqrt{Z}$ < 1.02 for most high-Tc superconductors. The symmetry rules, first proposed in [20] and observable in Table 1, are:

1. Materials with exactly the same average electronegativity, valence electrons and atomic number (case a) have the same Tc.

2. If two or more materials have the same average valence electrons Ne, and atomic number Z, (case b) then their Tcs will be proportional to their electronegativities.

3. If two or more materials have the same average electronegativity $\mathcal{X}$, and valence electrons Ne, (case c) then their Tcs will be proportional to their average atomic numbers, Z.

4. If two or more materials have the same average electronegativity $\mathcal{X}$, and average atomic numbers, Z (case d) then their Tcs will be proportional to their average valence electrons.

Using rule 1 above, we compare the MSCDs of 12 superconductors in Table 1 with the MSCDs of MgB$_2$ and LiMgN, shown in Table 2. The symmetry of their MSCDs gives us strong ground to predict that LiMgN will be a superconductor like MgB$_2$ but with Tc of 38.4K

## Iso-structural and Iso-valent Similarity

In the search for $MgB_2$-like superconductors, iso-structural and iso-electronic similarities were used for identifying likely candidates. Here we review a few examples. Using MSCD and the above symmetry rules, it will become clear why their Tcs are far from that of $MgB_2$. $MgB_2$ is hexagonal with $AlB_2$ type structure. Table 3 gives a list of compounds iso-structural and iso-electronic with $MgB_2$ and their Tcs. Table 4 gives the same compounds and their MSCDs. In Table 3, we have six sets of compounds out of nine sets that are both iso-structural and iso-valent with $MgB_2$. None of them has Tc anywhere near 39K. It is interesting to observe that none of them has the same electronegativity or atomic number as $MgB_2$. In Table 4, we show the MSCDs of the compounds. We find they fall into two classes: those with $Ne/\sqrt{Z}$ >1.0. There are four examples, and none of them is superconducting, as predicted by the symmetry rules [20]. $CaB_2$ was predicted to be a superconductor [6] with Tc higher than $MgB_2$. Estimating $CaB_2$ using the symmetry rules [20], it cannot have a higher Tc since its electronegativity and $Ne/\sqrt{Z}$ are less than those of $MgB_2$. Similarly the remaining examples in Table 4 have much lower electronegativities and $Ne/\sqrt{Z}$ than $MgB_2$. We observe a similarity of Ne and Z in $ZrZn_2$ and $SrGa_2$, suggesting similar properties. $ZrZn_2$ is a known ferromagnetic superconductor [29]. None of the thirteen compounds studied in Table 3 meets the symmetry criteria for high Tc like $MgB_2$, based on structure and electrovalent similarity alone.

## Discussion

The search for magnesium diboride-like superconductors has revealed that we still do not know all the parameters that control Tc [20]. Iso-structural and iso-valent similarity [3 - 16] did not yield Tc even close to 39K (see Tables 3 and 4). This suggests that these parameters alone were not sufficient in determining Tc. The search for decisive parameters led us to explore the correlations of electronegativity and atomic number and formula weight with superconductivity [20]. In that search we discovered the similarity rules with which we have been able to predict the occurrence of superconductivity from simple periodic table parameters without recourse to DFT calculations. The prediction of superconductivity in LiMgN (and many other compounds in reference [20] ) stands out as a strong test of the validity of those rules. Also we may note that nitrides show interesting superconductors and may have potential high-Tc superconductors [29] like the oxides. Many examples exist of semiconductors that superconduct [33, 34, 35] at low temperatures. It is known that ionic bonding also occurs even in the high temperature superconductors [32], with no rules forbidding it [30]. One of the implications of LiMgN being very similar to $MgB_2$ is that it may also be a two-gapped superconductor. The total photon energy distribution of fig. 5 in ref. 27 suggests that. Experimental tests will confirm or refute these predictions.

**Conclusion**
Even though structurally dissimilar, $MgB_2$ and LiMgN have the same electronegativity, valence electrons count, atomic number and almost the same formula weight (Table 2). LiMgN meets all the conditions [20] necessary and sufficient for two materials to have the same Tc. We conclude that the ordered phase ($\alpha$) of LiMgN in the cubic or orthorhombic structure will be found to be superconducting with Tc between 38K and 39K.

**Acknowledgements**
I acknowledge stimulating discussions with A.O.E. Animalu at University of Nigeria, Nsukka, in 2008 on LiMgN as a potential superconductor. Again, in 2011 I discussed with M.B. Maple and J.E. Hirsch at UC San Diego and also with M. J. Schaffer, then at General Atomics, San Diego and J.R. O'Brien at Quantum Design, San Diego. This research was supported by M. J. Schaffer.

## TABLES

Table 1: Cases of similar superconductors. Cases 1, 3 and 6 show that when the three parameters of $\mathcal{X}$, Ne and Z are almost the same for two or more materials, their Tcs are also almost the same. Adapted from Ref. [20].

Table 2: Material Specific Characterization Datasets (MSCDs) for $MgB_2$ and LiMgN. The exact match for $\mathcal{X}$, Ne and Z lends very strong grounds for predicting very close Tc as shown in six cases in Table 1.

Table 3: Some materials iso-structural and or iso-valent with $MgB_2$. Refrences are indicated.

Table 4: MSCD of materials iso-structural and iso-electronic or iso-valent with $MgB_2$. This table provides quantitative data for analysis of Tc, based on equation (1) in this paper.

| Superconductor | | $\mathcal{X}$ | Ne | Z | Ne/$\sqrt{Z}$ | Fw | Fw/Z | Tc(K) | Ko |
|---|---|---|---|---|---|---|---|---|---|
| 1 | NbN | 2.30 | 5.0 | 24 | 1.0206 | 106.913 | 4.45 | 17 | 7.55 |
| | MoC | 2.15 | 5.0 | 24 | 1.0206 | 107.951 | 4.50 | 14.3 | 6.79 |
| 2 | CaAlSi | 1.4333 | 3.0 | 15.667 | .7579 | 95.15 | 6.07 | 7.8 | 7.18 |
| | SrAlSi | 1.4333 | 3.0 | 21.667 | .6445 | 142.69 | 6.59 | 5.8 | 6.28 |
| 3 | ZrN | 2.2 | 4.5 | 23.5 | .9283 | 105.231 | 4.478 | 10.7 | 5.24 |
| | NbC | 2.05 | 4.5 | 23.5 | .9283 | 104.917 | 4.465 | 11.1 | 5.83 |
| 4 | Nb$_3$Sn | 1.65 | 4.75 | 43.25 | .7181 | 397.44 | 9.19 | 18 | 15.19 |
| | Nb$_3$Ge | 1.65 | 4.75 | 38.75 | .7631 | 351.34 | 9.07 | 23.2 | 18.43 |
| 5 | Nb$_3$Al | 1.575 | 4.5 | 34 | 0.7717 | 305.71 | 8.991 | 16-20 | 13.2—16.5 |
| | V$_3$Ga | 1.6 | 4.5 | 25 | 0.9 | 222.54 | 8.902 | 16.8 | 11.67 |
| 6 | BeB$_2$ | 1.8333 | 2.6667 | 4.6667 | 1.2344 | 30.63 | 6.56 | 0 | 0 |
| | LiBC | 1.8333 | 2.6667 | 4.6667 | 1.2344 | 29.96 | 6.42 | 0 | 0 |

Table 1: Cases of similar superconductors. Cases 1, 3 and 6 show that when the three parameters of $\mathcal{X}$, Ne and Z are almost the same for two or more materials, their Tcs are also almost the same. Adapted from Ref.[20]

| Material | | $\mathcal{X}$ | Ne | Z | Ne/$\sqrt{Z}$ | Fw | Fw/Z | Tc(K) | Ko |
|---|---|---|---|---|---|---|---|---|---|
| 1 | MgB$_2$ | 1.7333 | 2.667 | 7.333 | 0.9847 | 45.93 | 6.263 | 39 | 22.85 |
| 2 | LiMgN | 1.7333 | 2.667 | 7.333 | 0.9847 | 45.26 | 6.172 | 38.5 | 22.85 |

Table 2: MSCDs for MgB$_2$ and LiMgN. The exact match for $\mathcal{X}$, Ne and Z and close match of their Formula weights (Fw), lends very strong grounds for predicting very close Tcs of both materials, following the chemical symmetry rules of ref. [20].

| * = with MgB$_2$ (AlB2, C32, Hexagonal) | CaBeSi | LiBC | Li$_{0.5}$BC | BeB$_2$ | BeB$_{2.75}$ | CaGa$_2$, SrGa$_2$, BaGa$_2$ | CaAlSi, SrAlSi | CaB$_2$ | ZrZn$_2$ |
|---|---|---|---|---|---|---|---|---|---|
| Iso-structural * | Yes | Yes | No | Yes | No | Yes | Yes | Yes | Yes |
| Iso-valent* | Yes | Yes | No | Yes | No | Yes | No | Yes | **Yes** |
| Same Electronegativity* | **No** | **No** | **No** | **No** | **No** | **No** | **No** | **No** | No |
| Same Atomic Number* | No | No | No | No | No | No | No | No | No |
| Tc (K) | 0.4K | 0K | 0K | 0K | 0.7K | <1.02K | 7.8K, 5.8K | >39K?? | **<1.02K** |
| Reference | 5 | 7, 10 | 9 | 11 | 11 | 14 | 12, 14, 15 | 6 | 29 |

**Table 3: Materials iso-structural, iso-electronic and iso-valent with MgB$_2$. Iso-structural and iso-valent similarity did not influence upward value of Tc to near 39K. Table 4 explains why.**

| | Material | $\mathcal{X}$ | Ne | Z | Ne/$\sqrt{Z}$ | Fw | Fw/Z | Tc (K) | Ko | Ref | Comments |
|---|---|---|---|---|---|---|---|---|---|---|---|
| 1 | MgB$_2$ | 1.7333 | 2.6667 | 7.3333 | 0.9847 | 45.93 | 6.263 | 39 | 22.85 | 1, 16 | Ne/$\sqrt{Z}$ >1.0 |
| 2 | LiBC | 1.8333 | 2.6667 | 4.6667 | 1.2344 | 29.96 | 6.42 | 0 | 0 | 7, 10 | Ne/$\sqrt{Z}$ >1.0 |
| 3 | Li$_{0.5}$BC | 2.0 | 3.0 | 5.0 | 1.3416 | 26.29 | 5.26 | 0 | 0 | 9 | Ne/$\sqrt{Z}$ >1.0 |
| 4 | BeB$_2$ | 1.8333 | 2.6667 | 4.6667 | 1.2344 | 30.63 | 6.56 | 0 | 0 | 11 | Ne/$\sqrt{Z}$ >1.0 |
| 5 | BeB$_{2.75}$ | 1.8667 | 2.7333 | 4.7333 | 1.2563 | 38.74 | 8.19 | <0.7 | 0 | 11 | Ne/$\sqrt{Z}$ >1.0 |
| 6 | CaB$_2$ | 1.6667 | 2.6667 | 10.0 | 0.8433 | 61.70 | 6.17 | ? | ? | 6 | Tc likely to be <39K |
| 7 | CaAlSi | 1.4333 | 3.0 | 15.667 | .7579 | 95.15 | 6.07 | 7.8 | 7.18 | 12, 14, 15 | Ne/$\sqrt{Z}$ <0.8 and $\mathcal{X}$<1.733 |
| 8 | SrAlSi | 1.4333 | 3.0 | 21.667 | .6445 | 142.69 | 6.59 | 5.8 | 6.28 | 12,14, 15 | Ne/$\sqrt{Z}$ <0.8 and $\mathcal{X}$<1.733 |
| 9 | CaBeSi | 1.4333 | 2.6667 | 12.667 | 0.7493 | 77.18 | 6.093 | 0.4 | 0? | 5 | Ne/$\sqrt{Z}$ <0.8 and $\mathcal{X}$<1.733 |
| 10 | CaGa$_2$ | 1.4 | 2.6667 | 27.333 | 0.5101 | 179.52 | 6.57 | <1.02 | 1.4 | 14 | Ne/$\sqrt{Z}$ <0.8 and $\mathcal{X}$<1.733 |
| 11 | ZrZn$_2$ | 1.5333 | 2.6667 | 33.333 | 0.4619 | 222.0 | 6.66 | <1.02 | 1.4 | 29 | Ne/$\sqrt{Z}$ <0.8 and $\mathcal{X}$<1.733 |
| 12 | SrGa$_2$ | 1.4 | 2.6667 | 33.333 | 0.4619 | 227.06 | 6.81 | <1.02 | 1.4 | 14 | Ne/$\sqrt{Z}$ <0.8 and $\mathcal{X}$<1.733 |
| 13 | BaGa$_2$ | 1.3667 | 2.6667 | 39.333 | 0.4252 | 276.77 | 7.04 | <1.02 | 1.4 | 14 | Ne/$\sqrt{Z}$ <0.8 and $\mathcal{X}$<1.733 |

Table 4: MSCD of materials iso-structural and or iso-valent with MgB$_2$ and their Tcs. Note that Ne/$\sqrt{Z}$ plays a key role in determining superconductivity. So too does the value of $\mathcal{X}$, as shown in equation (1) derived in ref. [20].

.